\documentclass[twocolumn,preprintnumbers,pre,floatfix]{revtex4}

\usepackage{dcolumn}
\usepackage{bm}
\usepackage{graphicx}

\begin{document}

\preprint{APS/PRE}

\title{Direct observation of crystal nucleation and growth in a quasi-two-dimensional nonvibrating granular system}

\author{A. Escobar, F. Donado\footnote{fernando@uaeh.edu.mx}}
\affiliation{Instituto de Ciencias B\'asicas e Ingenier\'{\i}a de la Universidad Aut\'onoma del Estado de Hidalgo-AAMF,\\Pachuca 42184, Pachuca, M\'exico\\
e-mail:fernando@uaeh.edu.mx}
\author{R. E. Moctezuma\footnote{rosario@ifisica.uaslp.mx}}
\affiliation{CONACYT-Instituto de F\'isica ``Manuel Sandoval Vallarta'',
Universidad Aut\'onoma de San Luis Potos\'i, Alvaro Obreg\'on 64,
78000 San Luis Potos\'i, S.L.P., M\'exico}
\author{Eric R. Weeks\footnote{erweeks@emory.edu }}
\affiliation{Physics Department, Emory University, Atlanta, Georgia 30322 USA.}

\date{\today}

\begin{abstract}
We study a quasi-two-dimensional macroscopic system of magnetic spherical particles settled on a shallow concave dish under a temporally oscillating magnetic field. The system reaches a stationary state where the energy losses from collisions and friction with the concave dish surface are compensated by the continuous energy input coming from the oscillating magnetic field. Random particle motions show some similarities with the motions of atoms and molecules in a glass or a crystal-forming fluid. Because of the curvature of the surface, particles experience an additional force toward the center of the concave dish. When decreasing the magnetic field, the effective temperature is decreased and diffusive particle motion slows.  For slow cooling rates we observe crystallization, where the particles organize into a hexagonal lattice.  We study the birth of the crystalline nucleus and the subsequent growth of the crystal. Our observations support non-classical theories of crystal formation. Initially a dense amorphous aggregate of particles forms, and then in a second stage this aggregate rearranges internally to form the crystalline nucleus.  As the aggregate grows, the crystal grows in its interior. After a certain size, all the aggregated particles are part of the crystal and after that, crystal growth follows the classical theory for crystal growth.
\end{abstract}


\maketitle
\section{Introduction}

The solidification process of a fluid through a controlled cooling process is a fundamental issue from both a scientific and technological point of view.  Solidification can result in a glass, a crystal, or a heterogeneous system containing amorphous and crystalline phases  \cite{ediger96,debenedetti01,stevenson2011,sosso2016}. If the solidification process could be completely understood and controlled, it would allow us to make materials with specific properties.  Crystalline materials are used in countless technological applications due to their distinctive electrical, optical, and magnetic properties.  

A crystal is a solid phase with ordered structure that can be obtained from a liquid through a cooling process, or from an amorphous solid through an annealing process. Although there is currently much indirect information about the crystallization process, direct observation of the motion of individual atoms (``particles'') while a crystal is forming is challenging because the methods for resolving the particle size and the necessary temporal resolution have yet to be developed \cite{oxtoby2000}. Scattering techniques are used to study crystallization, however, the information we can obtain with this technique is incomplete. Therefore, complementary techniques should be used to deeply understand the crystallization process.

Classical nucleation theory describes homogeneous nucleation as due to spontaneous structural fluctuations which occasionally form ordered aggregates which then frequently dissolve back into the disordered liquid.  However, if an ordered aggregate is formed above a critical size, it will most likely grow to form a crystal \cite{sosso2016,zahn2015}.  The size where the probability of growth is equal to the probability of shrinking is termed the critical nucleus size, and is determined by the Gibbs free energy.  The change in bulk energy (negative) favors growing a crystal, and the change in the surface energy (positive) opposes growing a crystal.  At the critical size these two contributions to the Gibbs free energy are in balance; above the critical size the bulk energy reward for crystal growth dominates, stabilizing the aggregate and resulting in further growth.  Non-classical theories instead claim that the process involves at least two steps. In the first step, a disordered aggregate forms with some critical size and then in the second step the aggregate evolves into an ordered configuration to form a crystal nucleus \cite{sosso2016,zahn2015,gebauer2014}.  Research is in progress to give direct evidence in favor of classical nucleation or non-classical nucleation theory \cite{vekilov2010,deyoreo2013,gebauer2018}. 

Studies focused on a description at the particle level are key to support one or another theory. For instance, work has been done using proteins, where it is possible to study the crystallization phenomenon due to the large protein size compared to that of small molecules \cite{galkin1999,yau2000,vekilov2011}.  Experiments with colloids observed direct crystal nucleation \cite{gasser01,konig2005,assoud2009,wang2010} or two-step nucleation \cite{tan13}, although in the latter work it was unclear if the intermediate state was truly metastable or just a structural precursor as the sample structure continuously changed from disordered to ordered.  Overall, our understanding of the ways crystallization occurs in different systems is still limited.

The use of macroscopic model systems can help us to understand the crystallization mechanism because some systems allow a detailed description at particle level \cite{tsai2003,rietz2018,panaitescu10,panaitescu12,ebert2009,wang2010,konig2004,konig2005,sanchez19,reis2006,reis2007,daniels2005}. These systems exhibit different phases when a physical quantity such as volume fraction, viscosity, temperature, or particle concentration is varied. Under some particular conditions the formation of crystalline structures has been reached \cite{voronoi18,crystal20,cafiero2000}.  One kind of these macroscopic models are granular systems under mechanical vibrations. Their dynamics can be easily studied because their macroscopic particle motions are slow enough to be evaluated by using standard video techniques. The agitation of the system is induced by means of the container, which oscillates at a certain frequency, which allows control over particle dynamics \cite{blair2003,reis2006,daniels2005,reis2007,tsai2003,rietz2018}.  Also colloidal systems \cite{ebert2009,wang2012} have been used for the same purpose. In both cases, the inverse of the particle concentration acts as the control parameter mimicking the temperature; although for granular systems, the agitation also acts as a direct control parameter analogous to temperature \cite{blair2003,reis2007,morales2018}. 

In recent work we have studied a non-vibrating granular model for a fluid based on millimeter-sized magnetic balls under an oscillating magnetic field \cite{brownian17,crystal20}.  In these systems, the spheres have permanent magnetic dipoles, and the magnetic field oscillates vertically causing the spheres to roll to reorient their dipoles to match the field.  This random rolling motion causes the spheres to move nearly ballistically at short time scales and diffusively at longer time scales \cite{brownian17}.  Their velocity distribution follows a Maxwell-Boltzmann distribution that can be controlled by the amplitude of the applied magnetic field. From this distribution an effective temperature can be obtained \cite{cecilio16}. It was also found that sudden cooling leads the system to change from fluid-like to solid-like. This macroscopic model is ideal for studying solidification at the particle level, since it allows us to study the motion of individual particles at both short and long times \cite{voronoi18,crystal20}.

In Ref. \cite{crystal20} we settled the magnetic particles in a shallow concave dish where gravity enhances the concentration of particles in the center.  If the magnetic forcing is turned off quickly, the particles condense to a disordered aggregate under the gravitational influence.  However, much like molecular systems where the cooling rate matters, by decreasing the magnetic forcing very slowly, particles form crystalline structures.  The formation of different structures depending on the cooling rate were obtained:  glass (fastest quench rate), crystals (slowest quench rate), or mixed structures (intermediate quench rates).  For the slowest cooling rate the crystals are compact hexagonal arrangements. During these studies of the crystallization process, experimental evidence was found to favor a non-classical process for nucleation and crystal growth. However, this evidence was not studied in detail. 

In the present work, we are interested in the initial formation of the nucleus and how particles move to their final positions in a crystal. In particular, show evidence that nucleation takes place via a two-stage process:  first a disordered dense aggregate, and then a more ordered and more dense crystal nucleus.  We analyze and compare different structural characteristics of the system and determine their relationship with the stability of the aggregates. 

\section{Experimental setup}

\begin{figure}
\centering
\includegraphics[width=0.45\textwidth]{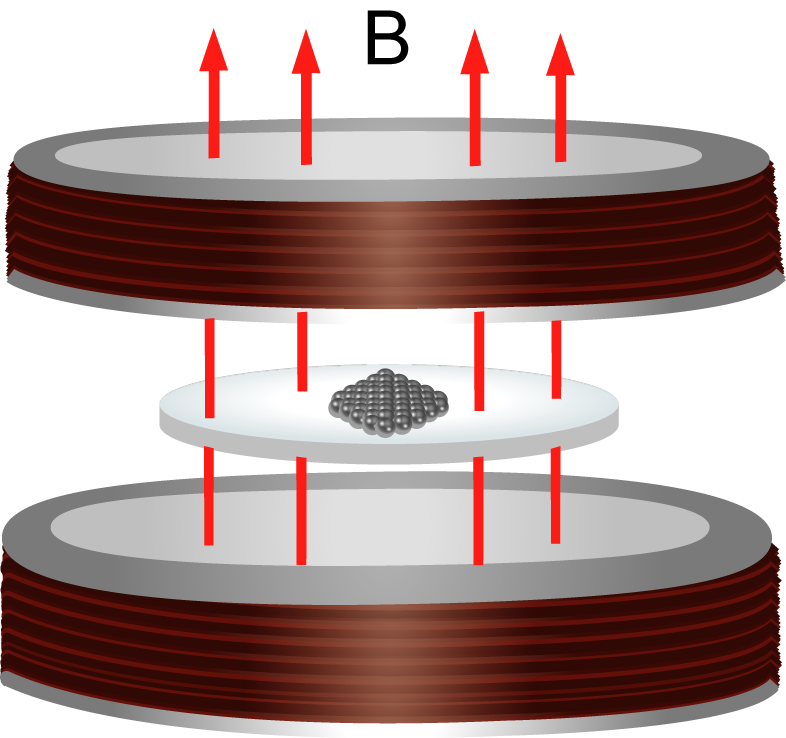}
\caption{ Experimental setup.  The Helmholtz coils have an inner diameter of 15~cm and an outer diameter of 20~cm. } \label{setup}
\end{figure}

Particles are settled on a concave lens of -250 mm focus length and 50.8 mm in diameter.  The lens is located in the middle of a pair of Helmholtz coils (Fig.~\ref{setup}) which produce a vertical magnetic field fed by a Kepco BOP 36-6 M power amplifier controlled by a PC through a DAQ card, using LabView.  The particles are 131 steel balls of 1 mm of diameter, ANSI 420 grade 1000 by Gimex S.A.  The experiments are recorded using a CCD camera at 30 fps in AVI interlaced format.  To improve the visual definition of particles centers and increase the time resolution, we use a deinterlace filter obtaining a final 60 fps resolution.  We use ImageJ and its plugin Mosaic to follow particle trajectories \cite{imagej,mosaic}. Our spatial resolution is $\approx$0.07 mm ($x$ and $y$ position).

The system is subject to oscillatory magnetic field of the form $B_o=A(t) \sin(2 \pi f t)$.  The experiment starts with the amplitude $A(t)=66$~G and changes following a decreasing stepladder function at 0.02 G/s.  The frequency $f$ is kept constant at 9.24 Hz. The oscillating magnetic field causes particles to rotate to follow the magnetic field direction. Because of the friction of particles with the base of the container, particles roll as they rotate.  However, they cannot fully align their permanent magnetic dipoles with the imposed field before the field reverses direction, therefore,  the rolling motion is erratic and particles frequently change their directions.  For large $A(t)$ the particle motion is essentially diffusive except at very short time scales where it is almost ballistic; at these shortest time scales we observe the mean square displacement (MSD) growing as $\sim  \Delta t^{1.7}$ \cite{tapia20}.  Since the magnetic field plays the role of the temperature, henceforth, we will refer to the amplitude of the magnetic field as the effective temperature \cite{cecilio16}. In this sense, we model a cooling rate by gradually decreasing the amplitude of the magnetic field.  As the magnetic field decreases, particle motions become subdiffusive, and eventually at low enough magnetic field the particles are arrested.  In the next sections we describe the structural and dynamical changes during this cooling process.

\section{Experimental results}

\subsection{Structural analysis considering all particles}
\label{systemaswhole}

\begin{figure}
\centering
\includegraphics[width=0.45\textwidth]{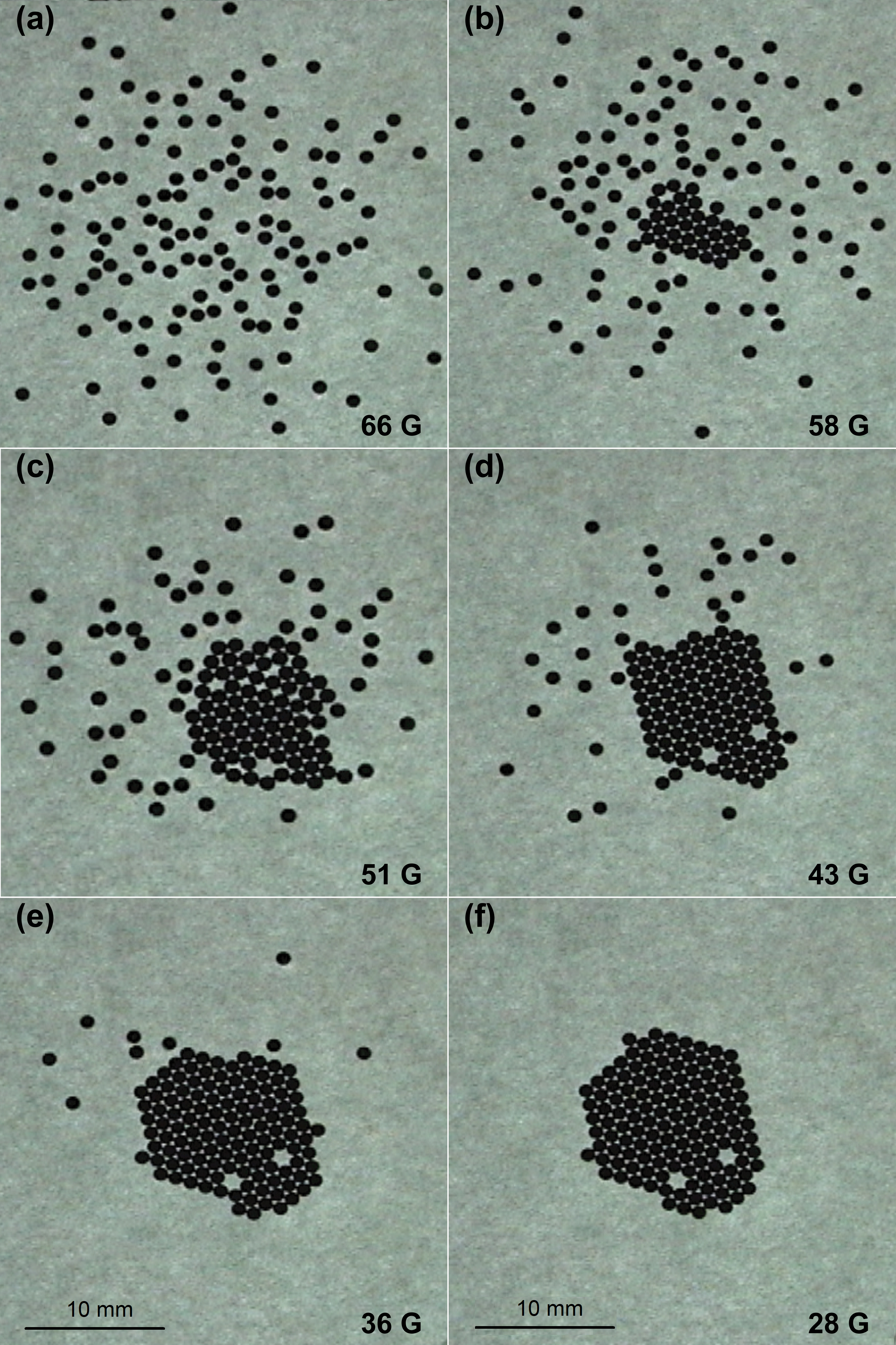}
\caption{Sequence of photos from experiment E1, showing evolution from a fluidlike configuration to crystal configuration.  The amplitude of the forcing $B_o$ is noted in the corner of each image.} \label{sequence}
\end{figure}

In this section we present results considering all the particles in the system, both those in crystalline regions and those in amorphous regions.  Given that the nucleation process is random, we analyze three experiments of crystallization under identical conditions; these are referred to as E1, E2 and E3.  Figure \ref{sequence} shows a sequence of photos from E1. The system started in a disordered configuration where all the particles are separated from each other in a gas-like configuration.  Spontaneous particle-concentration fluctuations drive the formation of small aggregates of different sizes. We considered that particles form an aggregate when they are in contact more than our resolution time ($>1/60$ s). Small aggregates, below 4 particles, are very unstable.  As the magnetic field is slowly decreased, we observe the formation of a nucleus in Fig.~\ref{sequence}(b) and a subsequent growth of the crystal as the field decreases further.  The aggregate is stabilized by friction:  when particles touch each other, they experience frictional contact.  These interactions can prevent them from rolling when the magnetic field direction changes, thus, the more neighbors a particle has in an aggregate, the more frictional contacts stabilize the particle.  Nonetheless, particles at the boundary of an aggregate also experience random kicks from colliding gaseous particles, which can destabilize them and cause the boundary particles to ``evaporate'' into the gas.  It is the competition between the frictional stabilization and the random forces that determine the possibilities of nucleating and growing an aggregate.  Of course, a hexagonally ordered aggregate allows the interior particles to have a maximum number of frictional ``bonding contacts'' $N_B = 6$ and thus should be maximally stable.

\begin{figure}
\centering
\includegraphics[width=0.45\textwidth]{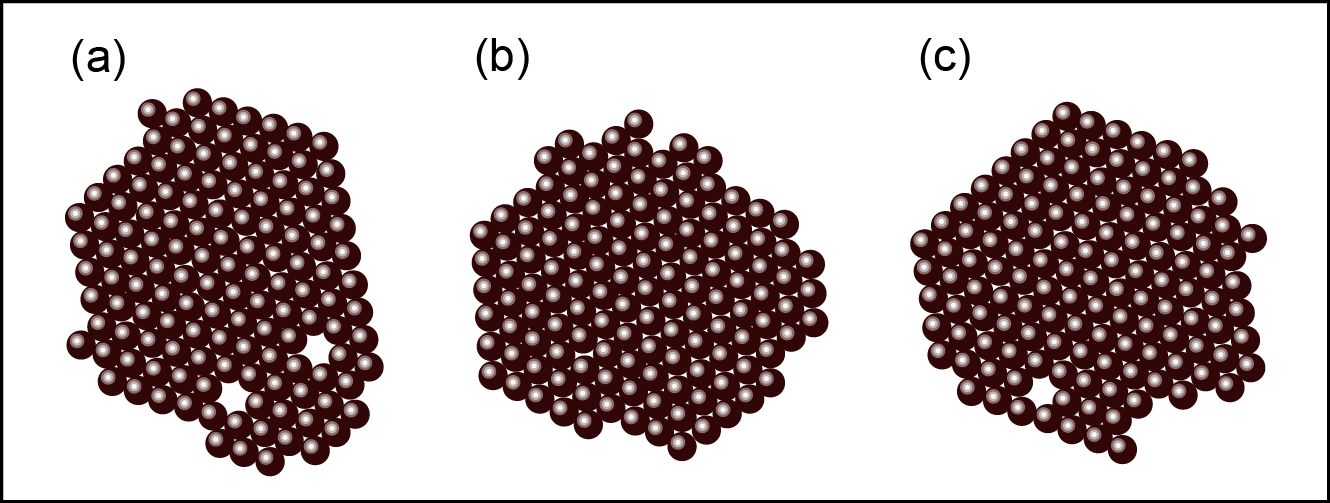}
\caption{Final configurations of experiments E1, E2, and E3.  Each particle is colored according to the sixth bond configurational order parameter $\psi_{6}^{'}$. }
\label{phi6final}
\end{figure}

\begin{figure}
\centering
\includegraphics[width=0.45\textwidth]{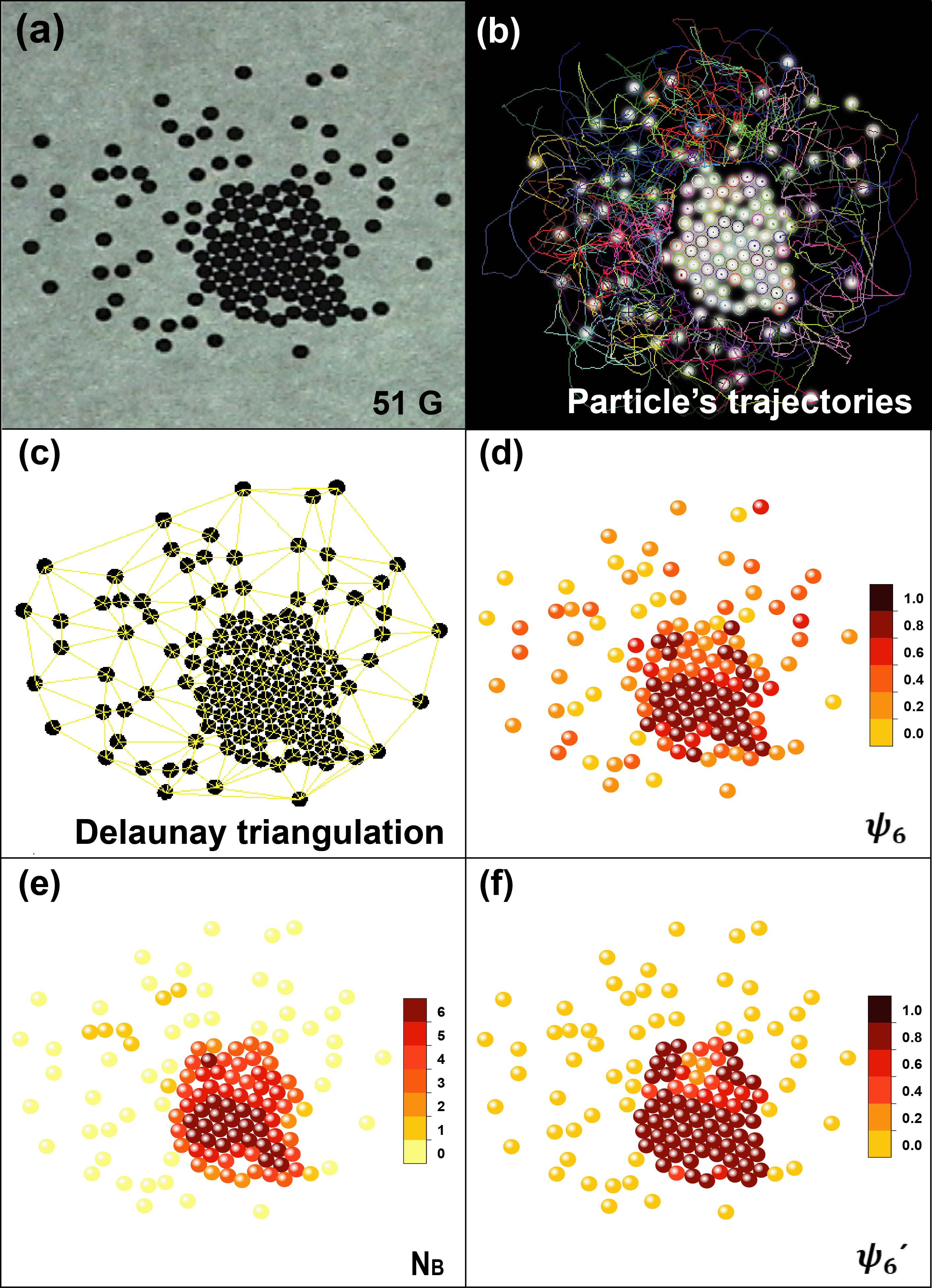}
\caption{(a) A typical photo at a certain stage of the experiment, (b) trajectories of particles over 1.66~s (0.83~s before and 0.83~s after the position shown),
(c) Delaunay triangulation, (d)  $\psi_6$ order parameter,
(e) number of bonded neighbors $N_B$, and (f) the modified order parameter $\psi_{6}^{'}$, which is calculated based only on touching neighbors (those counted by $N_B$).
This configuration is the same shown in Fig.~\ref{sequence}(c).}
\label{parameters}
\end{figure}

Final crystalline states of the three experiments are shown in Fig.~\ref{phi6final}, confirming the hexagonal ordering.  We observe faceted boundaries, and for two configurations we see hexagonal vacancies.  We wish to characterize the crystallization process by examining the structure of the entire system as a function of time (and thus as a function of decreasing magnetic field).  Quantities of interest are shown in Fig.~\ref{parameters} and will be described next, measured from snapshots analyzed every 1.66~s to find signatures of crystallization.

\begin{figure}[!tbp]
\centering
\begin{center}
\leavevmode 
\includegraphics[width=8cm]{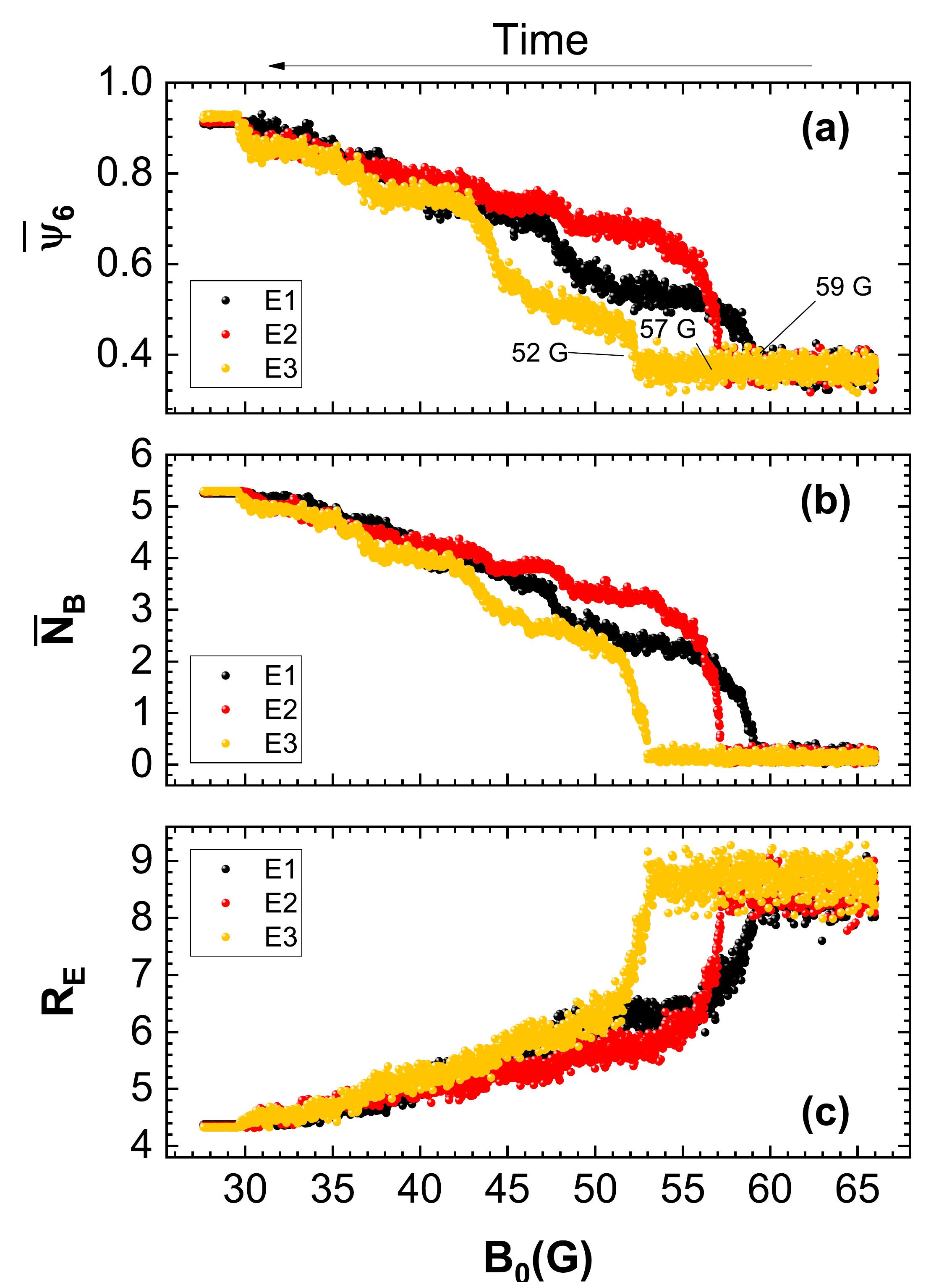}
\caption{\small (a) Average of the configurational order parameter $\psi_6$ for each frame as a function of magnetic field.
(b) Average of the number of bonded neighbors of particles for each frame as function of magnetic field.
(c) Effective system radius in each frame as a function of magnetic field, in units of the particle diameter $\sigma$.
}
\label{systemvstime}
\end{center}
\end{figure}

We start by identifying topological neighbors using a Delaunay triangulation of the particle configurations, shown in Fig.~\ref{parameters}(c).  From these topological neighbors we calculate the orientational order parameter $\psi_6$ for each particle $i$, defined as
\begin{equation}
\psi_6= \frac{1}{N_i}\sum_j\exp(6\text{i}\theta_{j}),
\end{equation}
where the sum on $j$ is over the $N_i$ neighbors of this particle and $\theta_{j}$ is the angle formed between the $x$-axis and the vector pointing from $i$ to $j$.  $\psi_6$ is a complex number and  we take the magnitude to quantify hexagonal order.  A particle in an hexagonally ordered region has $\psi_6=1$, and $\psi_6$ can be as low as zero for a particle in a disordered region; see Fig.~\ref{parameters}(d).  We plot the particle-averaged $\bar{\psi_6}$ as a function of magnetic field in Fig.~\ref{systemvstime}(a).  Going from right to left in the graph we see that the system starts in a disordered configuration ($\bar{\psi_6} \approx 0.36$)  that becomes ordered as the magnetic field decreases. The value of $\bar{\psi_6}$  clearly starts to increase at a certain value of the magnetic field that is different in each experiment. 

\begin{figure}[!tbp]
\centering
\begin{center}
\leavevmode 
\includegraphics[width=7cm]{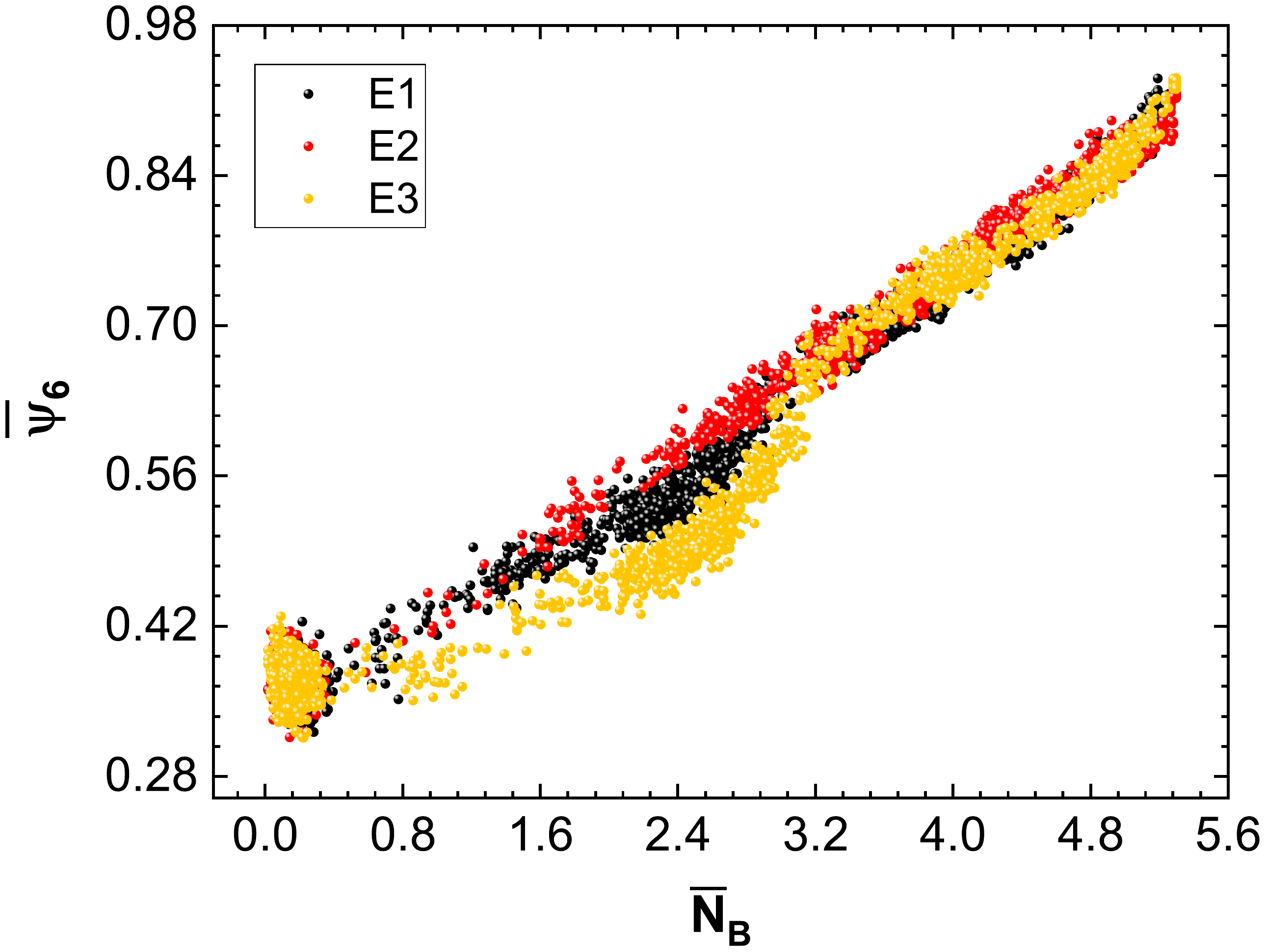}
\caption{\small Mean order parameter $\bar{\psi_6}$ as function of the mean bond number.  Each point represents a different time in the experiment, and the averages are over all particles at that time.}
\label{comparbondphi}
\end{center}
\end{figure}

An increasing $\bar{\psi_6}$ does not necessarily require particles to be in a dense aggregate.  To investigate this, we next determine the number of ``bonds'' $N_B$ for each particle.  $N_B$ is the number of neighbors that are in contact with a given particle, defined as center-to-center separations of less than $1.1\sigma$ in terms of the particle diameter $\sigma$; see Fig.~\ref{parameters}(e).  Figure \ref{systemvstime}(b) shows the average of $N_B$ as a function of the magnetic field.  This quantity follows a similar behavior as the one followed by the average of $\psi_{6}$, namely a low value (close to zero) when the particles are in a dilute gas-like state, and then a sharp increase as the initial aggregate forms.  Figure \ref{comparbondphi} shows the relation between the mean value of $N_B$ and the mean value of $\psi_6$.  The gas-like state corresponds to the lower left corner of this plot, and the initial jump in $\bar{N_B}$ is followed by growth of both $\bar{N_B}$ and $\bar{\psi_6}$.  This growth process shows an almost linear relation between these two quantities, albeit with some variability between the three experiments during the initial aggregation period.

We next examine the overall state of the system, both aggregate and surrounding particles, by calculating the effective system radius $R_E$.  This is the mean distance between each particle in the system and the system's instantaneous center of mass.  Figure \ref{systemvstime}(c) shows how this quantity starts at a large initial value, when the system is behaving more like a gas (at high magnetic forcing); $R_E$ then decreases abruptly at the formation of the first aggregate, and continues to decrease further as the magnetic field decreases.  Figure \ref{sequence} shows that as the aggregate is formed, the gas phase surrounding it moves in closer to the aggregate.  Of course, $R_E$ also decreases simply because the aggregate has many particles close to the system center of mass.

\begin{figure}[!tbp]
\centering
\begin{center}
\leavevmode 
\includegraphics[width=8cm]{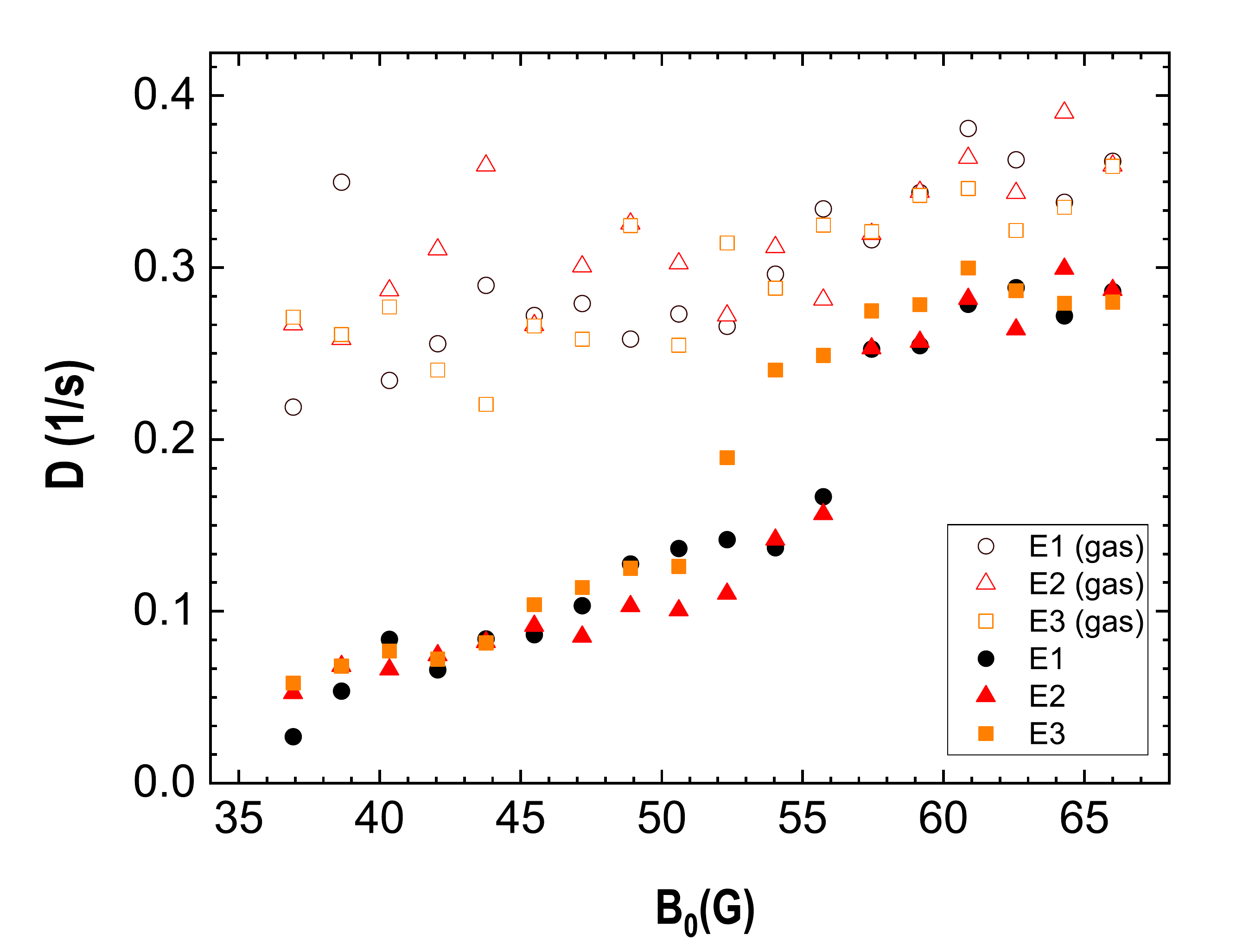}
\caption{\small  Effective diffusion coefficient measured in time windows of 4.16 s every 166.66 s.  The open symbols correspond to gaseous particles, and the filled symbols are averaged over all particles (both gaseous and aggregated).  The units are $\sigma^2/s$ in terms of the particle diameter $\sigma$.}
\label{dvstime3exp}
\end{center}
\end{figure}

To finish our description of the system as a whole during the quenching process, we analyze the particle trajectories such as those shown in Fig.~\ref{parameters}(b) to determine the effective diffusion coefficient at different times.  The data are shown in Fig.~\ref{dvstime3exp}, and show a sharp decrease when the initial stable aggregate forms.  The drop is due to the average of the gaseous particles (which stay fairly diffusive, see the open symbols in the figure) and the aggregated particles (which are essentially motionless, although at times exchanging with the gaseous particles).

The initial rapid changes seen in Fig.~\ref{systemvstime} occur as the magnetic field drops by $\sim 1$~G.  This occurs over 50~s.  Based on the typical diffusivity $D \approx 0.25$~(1/s) of the gas particles right before the initial aggregation event, we estimate that the particles' mean square displacement over this time interval is $\langle \Delta r^2 \rangle = 4 D \delta t = 50 \sigma^2$.  This shows that the gas particles are able to explore large distances during the initial aggregation event, implying that the initial growth is unlikely to be diffusion limited.

We additionally note that these figures show a systematic difference between the experiments:  the first experiment (E1) nucleates a stable aggregate earliest at the highest magnetic forcing $B_0$, while the third experiment (E3) nucleates the stable aggregate latest. This is due to the gradual increasing magnetization of the particles over the course of the project \cite{cecilio16}.  Further evidence of this increasing magnetization is in the initial plateau height of the $R_E$ data of Fig.~\ref{systemvstime}(c), which is largest for E3.  In this situation, the more strongly magnetized particles respond more forcefully to the oscillating external magnetic field, causing higher kinetic energy and thus a higher effective internal pressure, leading to the larger $R_E$ values.

\subsection{Structural analysis considering particles in aggregate}

The analysis in the previous subsection considered all of the particles in the experiment.  We now turn from the global to the local;  we wish to understand the stable aggregate once it forms and grows.  In each experiment, only one stable aggregate forms.  It grows as the forcing magnetic field is decreased until all of the particles belong to the aggregate.  To examine the growth of this aggregate, we analyze our data at 1.66~s intervals.  Aggregates are defined based on touching particles (ones with center-to-center separation less than 1.1$\sigma$ as mentioned in the previous section).  We discard aggregates smaller than 4 particles, as they typically are stable for less than 1~s.  At times when the stable aggregate had formed, we only found a few rare cases where there was more than one aggregate present in the image; and in all cases the stable aggregate is the biggest aggregate.

\begin{figure}[!tbp]
\centering
\begin{center}
\leavevmode 
\includegraphics[width=8cm]{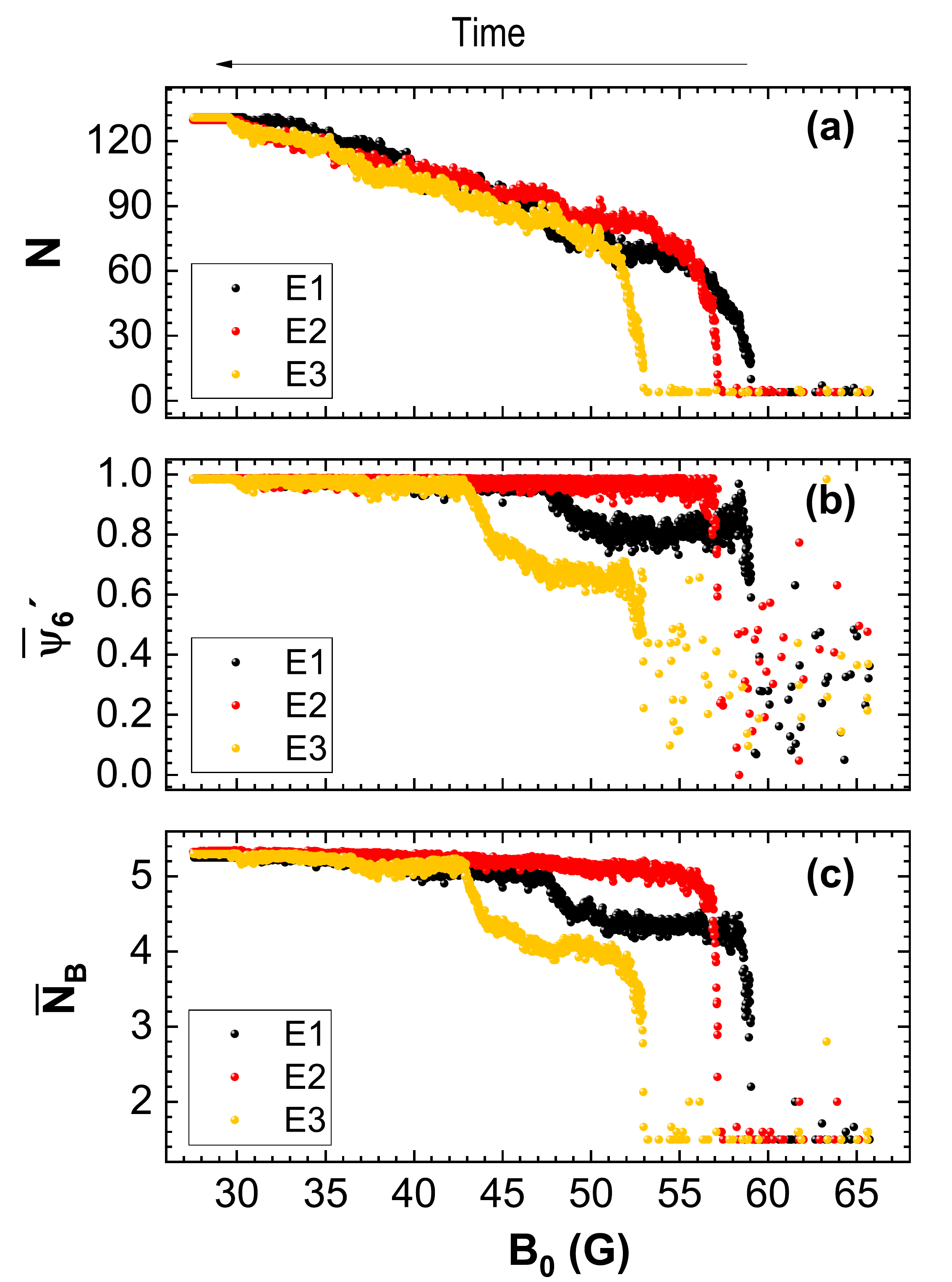}
\caption{\small 
(a) Aggregate size (number of particles $N$) as a function of the magnetic field. 
(b) Mean $\psi_{6}^{'}$ order parameter for the aggregate particles as a function of the magnetic field.
(c) Mean number of bonded neighbors $N_B$ for the aggregate particles as a function of the magnetic field.}
\label{aggregatevstime}
\end{center}
\end{figure}

Figure \ref{aggregatevstime}(a) shows the size of the aggregate as a function of the magnetic field.  At early times (large $B_o$) we observe small unstable aggregates that are made up of about four particles.  At some point an aggregate stabilizes and then begins to grow; roughly speaking, Fig.~\ref{aggregatevstime}(a) shows that when $N \gtrsim 10$ the aggregate grows irreversibly.  We wish to correlate the growth in size with the increase in ordering.  The hexagonal order parameter $\psi_6$ is based on neighboring particles defined by the Delaunay triangulation.  This is less useful for the aggregate, as particles at the edge of the aggregate have Delaunay neighbors that are not in the aggregate and not expected to be ordered.  Accordingly, we define a modified order parameter $\psi_{6}^{'}$ based only on the $N_B$ particles in contact with a given particle.  The difference in $\psi_{6}$ when considering all the particles in the system and when considering only the particles that form an aggregate can be seen in Figs.~\ref{parameters}(d) and (f).  Note that all the particles in Fig.~\ref{phi6final} have $\psi_{6}^{'}=1$.

In Fig.~\ref{aggregatevstime}(b) we show $\psi_{6}^{'}$ averaged over all aggregated particles.  At the largest magnetic field (earliest times), prior to the formation of a stable aggregate, it is observed that the aggregates usually are linear aggregates with $\psi_{6}^{'} \lesssim 0.4$. After the stable aggregate is formed, the size and the hexagonal ordering of the aggregate increased quickly to $0.6 \leq \psi_{6}^{'} \leq 0.9$.  For two of the experiments shown in Fig.~\ref{aggregatevstime}(b), this intermediate state is stable for a range of magnetic forcing.  Subsequently after a period of reordering, the aggregate shows nearly ideal hexagonal ordering ($\psi_{6}^{'} \approx 1.0$) and continues growing in that way.  A final view of the aggregate growth is depicted in Fig.~\ref{aggregatevstime}(c), where the number of neighbors $N_B$ a particle has within the aggregate is shown.  Again, two experiments show a plateau with $N_B \approx 4$ before final growth to $N_B > 5$.  (The maximum value of $N_B$ is 6 for particles in the interior of the aggregate, but because of the particles on the boundary with fewer neighbors, the mean value for $N_B$ does not reach 6.)  The results shown in Fig.~\ref{aggregatevstime}(b,c) show that after reaching a certain size the nucleus has the same ordered structure as the final crystalline phase, in accordance to classic nucleation theory.  Prior to this point, the growth is nonclassical as the initial metastable aggregate is not well-ordered.  This demonstrates two stages of growth of the crystal.  We note that experiment E2 (the red symbols in Fig.~\ref{aggregatevstime}) show an aggregate that appears to bypass the intermediate state, or at least to not linger in the intermediate state.

\begin{figure}[!tbp]
\centering
\begin{center}
\leavevmode 
\includegraphics[width=8cm]{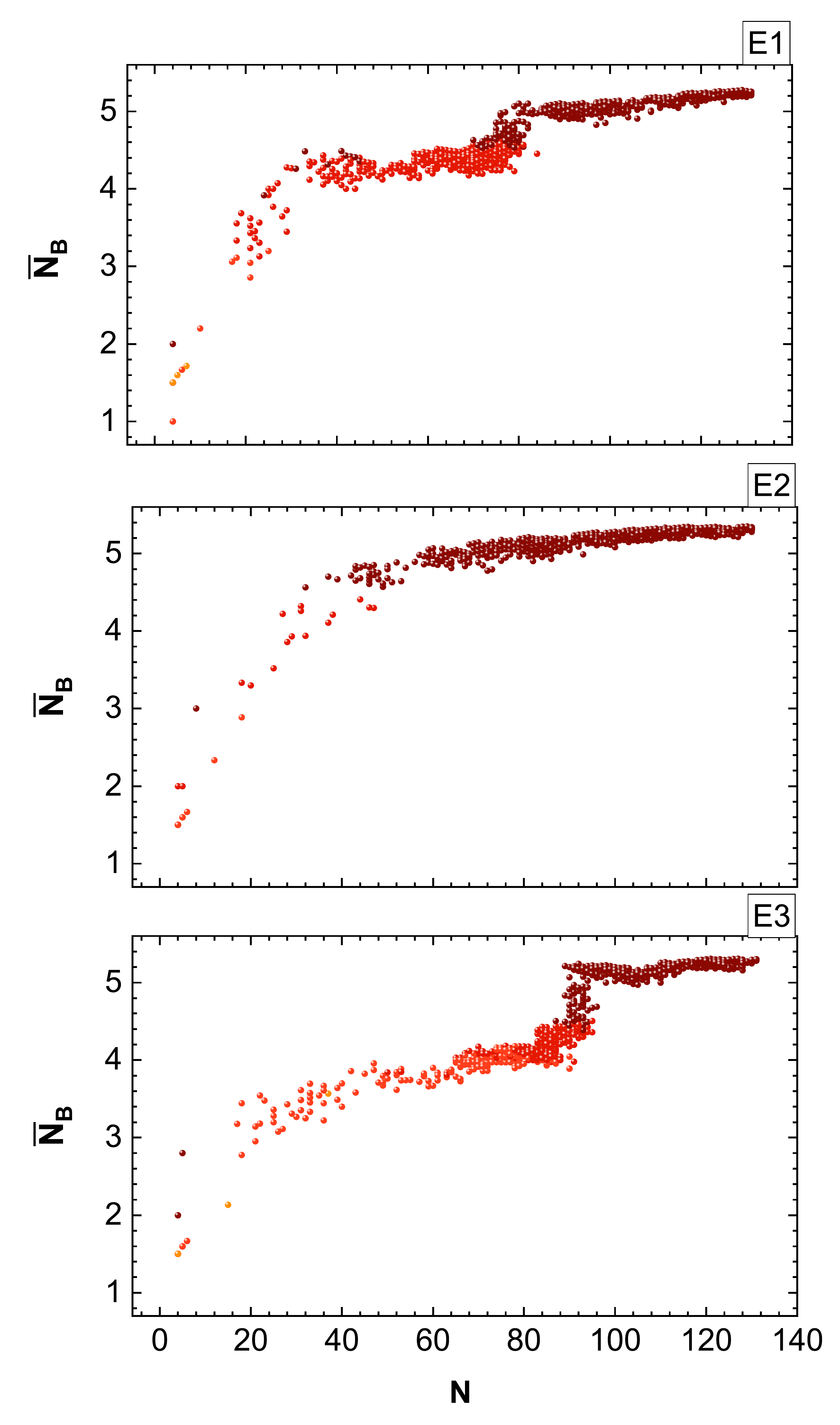}
\caption{\small  Average number of bonds $N_B$ as function of the number of particles $N$ in the aggregate.  The color indicates the mean value of $\psi_{6}^{'}$, matching the color key of Fig.~\ref{parameters}(f). }
\label{comparpartbondphi}
\end{center}
\end{figure}

\begin{figure}[!tbp]
\centering
\begin{center}
\leavevmode 
\includegraphics[width=7cm]{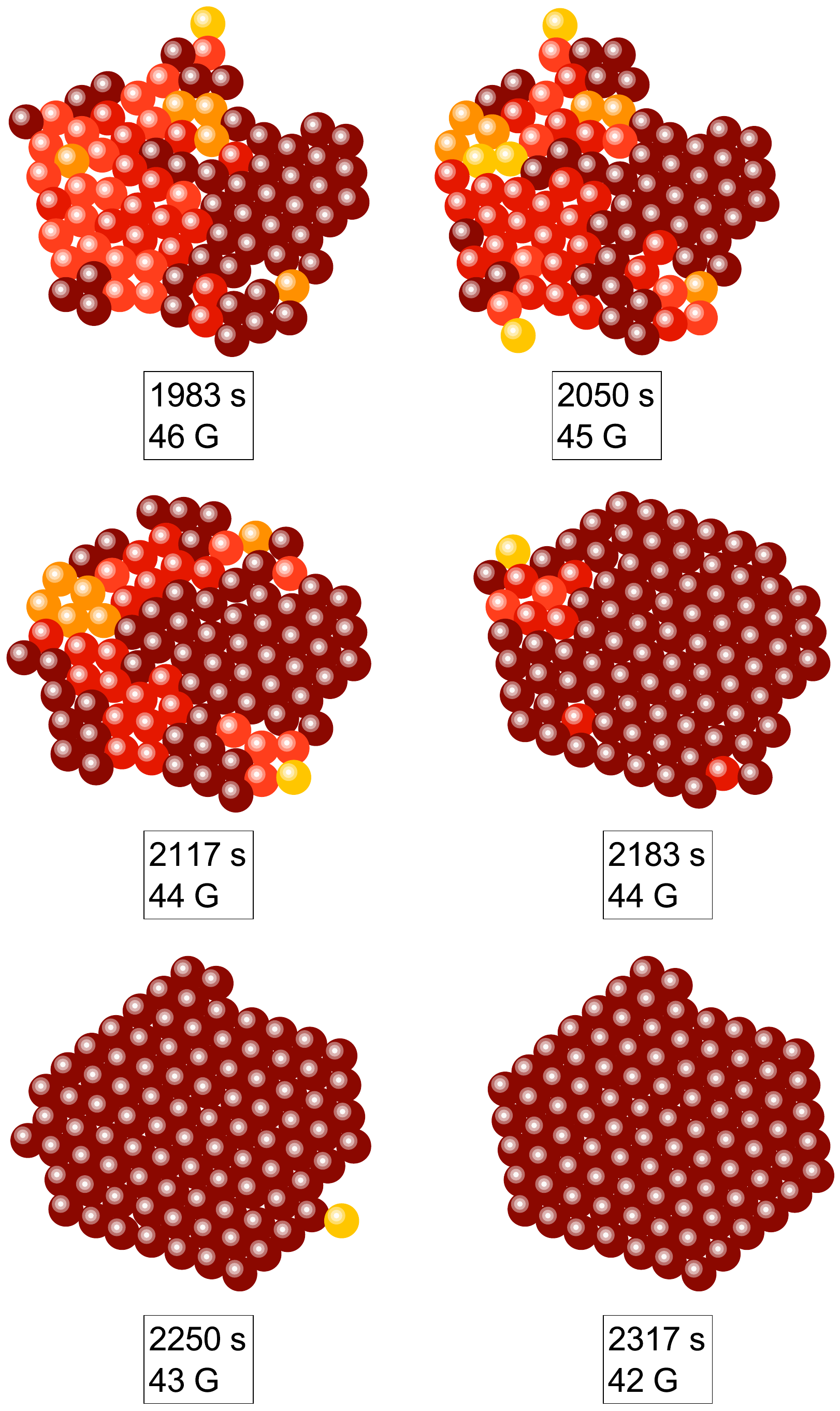}
\caption{\small  Sequence of images showing a notable change of structural characteristics.  The data are from experiment E3, and the color indicates $\psi_{6}^{'}$ of each particle, matching the color key of Fig.~\ref{parameters}(f).}
\label{graphs}
\end{center}
\end{figure}

In Fig.~\ref{comparpartbondphi} we observe the relation between the number of particles $N$ in the aggregate and the mean number of bonds $N_B$ each particle has. At the beginning, both quantities grow quickly.  There is then slow growth of $N_B$ for $N$ roughly between 40 to 80 particles.  In two experiments, E1 and E3, there is a sudden growth of $N_B$ coincident with a sudden growth in $\psi_{6}^{'}$ (indicated by the color change in Fig.~\ref{comparpartbondphi}).  After the transition the aggregate is ordered.  Figure \ref{graphs} shows an example of this change from a disordered aggregate to an ordered aggregate, taking place over 335~s.  This sequence correspond to the jump in Fig. \ref{comparpartbondphi}(c) at $N\approx 90$. 

\begin{figure}[!tbp]
\centering
\begin{center}
\leavevmode 
\includegraphics[width=7cm]{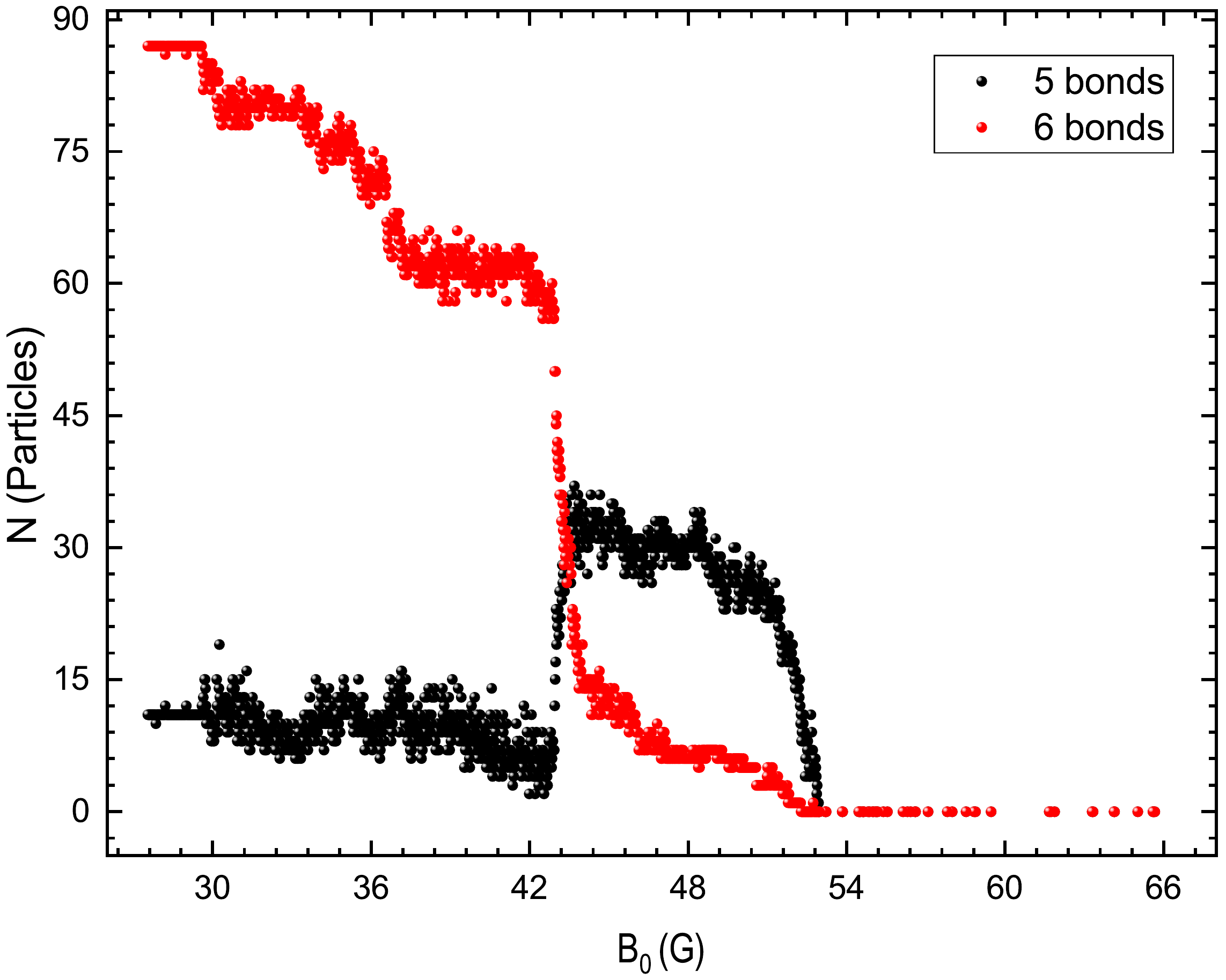}
\caption{\small  Number of particles with 5 and 6 bonds as a function of the magnetic field for the aggregate from experiment E3. }
\label{5and6}
\end{center}
\end{figure}

Another way to quantify the two-step crystallization process is to consider the number of particles in the aggregate with exactly $N_B = 5$ or 6 neighbors; this is shown in Fig.~\ref{5and6}.  There is a period of time for which many particles have $N_B=5$, followed by a rapid rearrangement so that many particles switch to having $N_B=6$ neighbors.  This corresponds to the increase in hexagonal order shown in Fig.~\ref{graphs}.

\subsection{Initial formation of the nucleus}

We have showed in the above section, that crystal formation started with a disordered aggregate which evolves toward an ordered aggregate containing all particles. To determine in a more precise way the initial formation of a nucleus we analyze in detail, frame by frame, the videos of the formation of the crystal. We focus our attention on the period from the formation of a stable aggregate to the formation of the first ordered structure with a hexagonal arrangement within the stable aggregate.  

\begin{figure}[!tbp]
\centering
\begin{center}
\leavevmode 
\includegraphics[width=8cm]{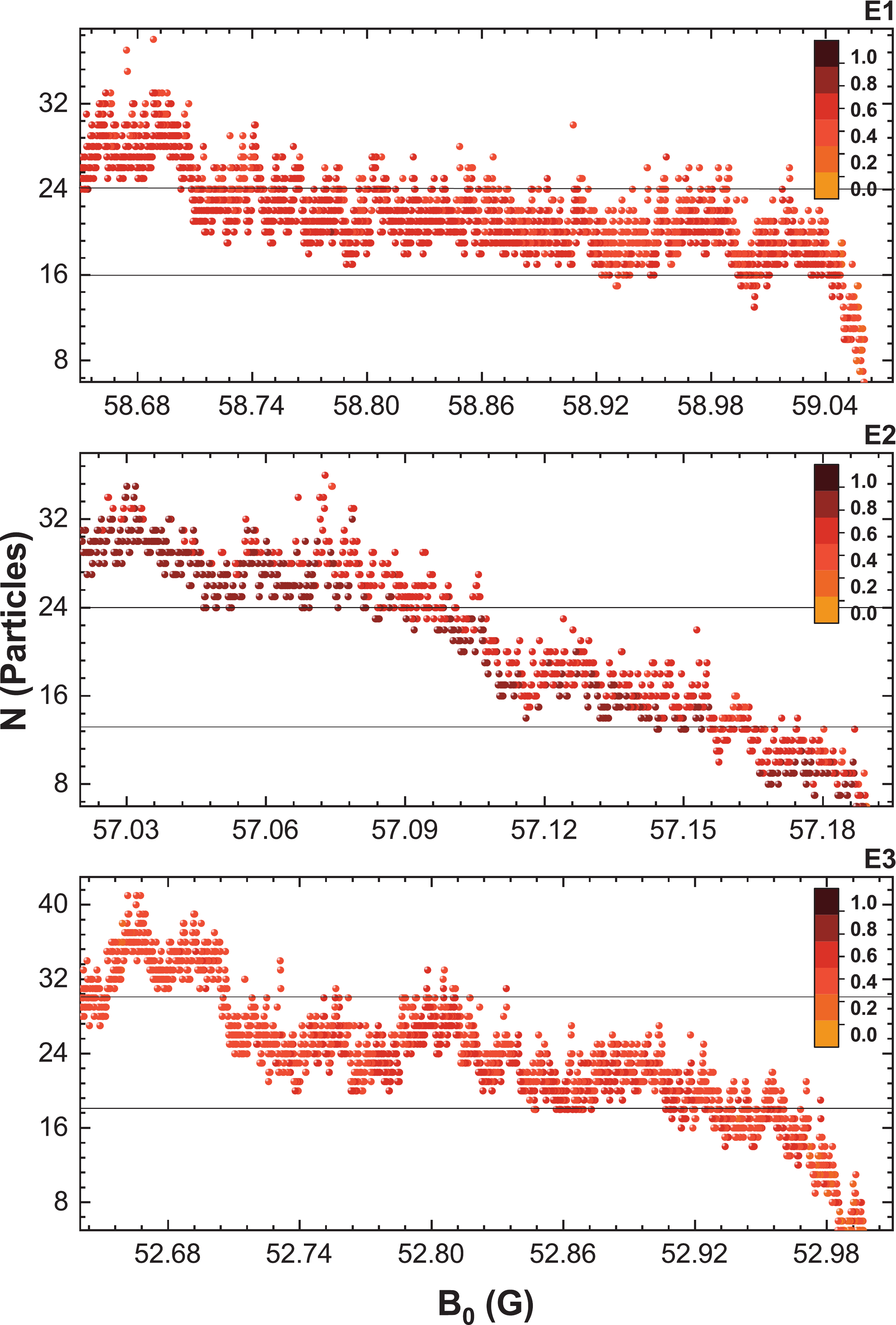}
\caption{\small Number of particles in the aggregate as function of the magnetic field amplitude with a time resolution of 1/60 s.  The horizontal lines separate out the early stage, middle stage, and late stage of growth.  The color indicates the value of $\psi_{6}^{'}$ as indicated in the legend.}
\label{npartphi6}
\end{center}
\end{figure}

\begin{figure}
\centering
\includegraphics[width=0.45\textwidth]{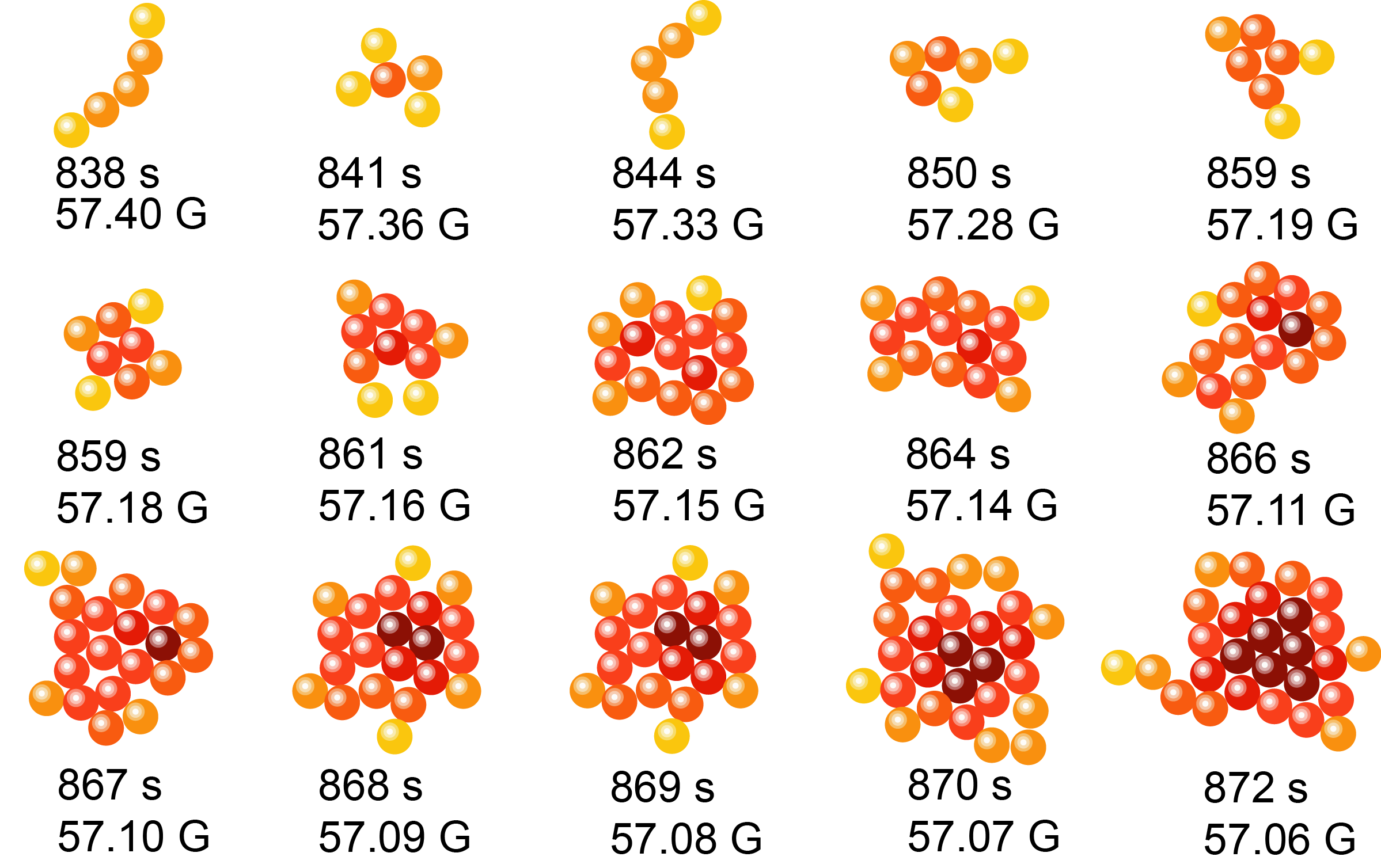}
\caption{\small  Sequence of particle configurations in the early formation of a nucleus, corresponding to the data in Fig.~\ref{npartphi6}(b).  The earliest stage has $N<16$ particles in the aggregate and corresponds to $B_o \geq 57.16$.  The second stage has $16 \leq N \leq 24$ and corresponds to $57.10 \leq B_o < 57.16$.  The third stage has $N>24$ and $B_o < 57.10$.  The color of each particle indicates $\psi_{6}^{'}$ matching the color key of Fig.~\ref{parameters}(f).  }
\label{threesteps}
\end{figure}

Figure \ref{npartphi6} shows in detail the growth of the initial stable aggregate, where the color indicates the mean value of the local hexagonal order parameter $\psi_{6}^{'}$.  
The initial formation of a stable aggregate takes some time to occur, but once it forms it quickly grows (right side of the plots in Fig.~\ref{npartphi6}, data below the lower horizontal lines).  Usually, the first stable aggregate formed is a ring-shape aggregate. It is stable in the sense it was not destroyed although it changes its form to be more compact.  After that, each experiment shows a rough plateau in the number of particles $N$ composing the aggregate; these are the points between the pairs of horizontal lines in Fig.~\ref{npartphi6}.   Usually in this stage the aggregate has substructures with hexagonal order. Then, there is a third stage as the aggregate again grows in size (left side of the plots in Fig.~\ref{npartphi6}) where more substructures with hexagonal order appear.  Figure \ref{threesteps} shows representative particle configurations of the aggregate in each of the stages. In the second stage it is observed the formation of substructures with some hexagonal order. In the third stage we see rearrangements leading to substantial hexagonal order in the interior of the aggregate.  While the aggregate still has disordered regions, the hexagonal ordering is essentially monotonically increasing at this point [in agreement with Fig.~\ref{aggregatevstime}(b)].
  
\begin{figure}[!tbp]
\centering
\begin{center}
\leavevmode 
\includegraphics[width=7cm]{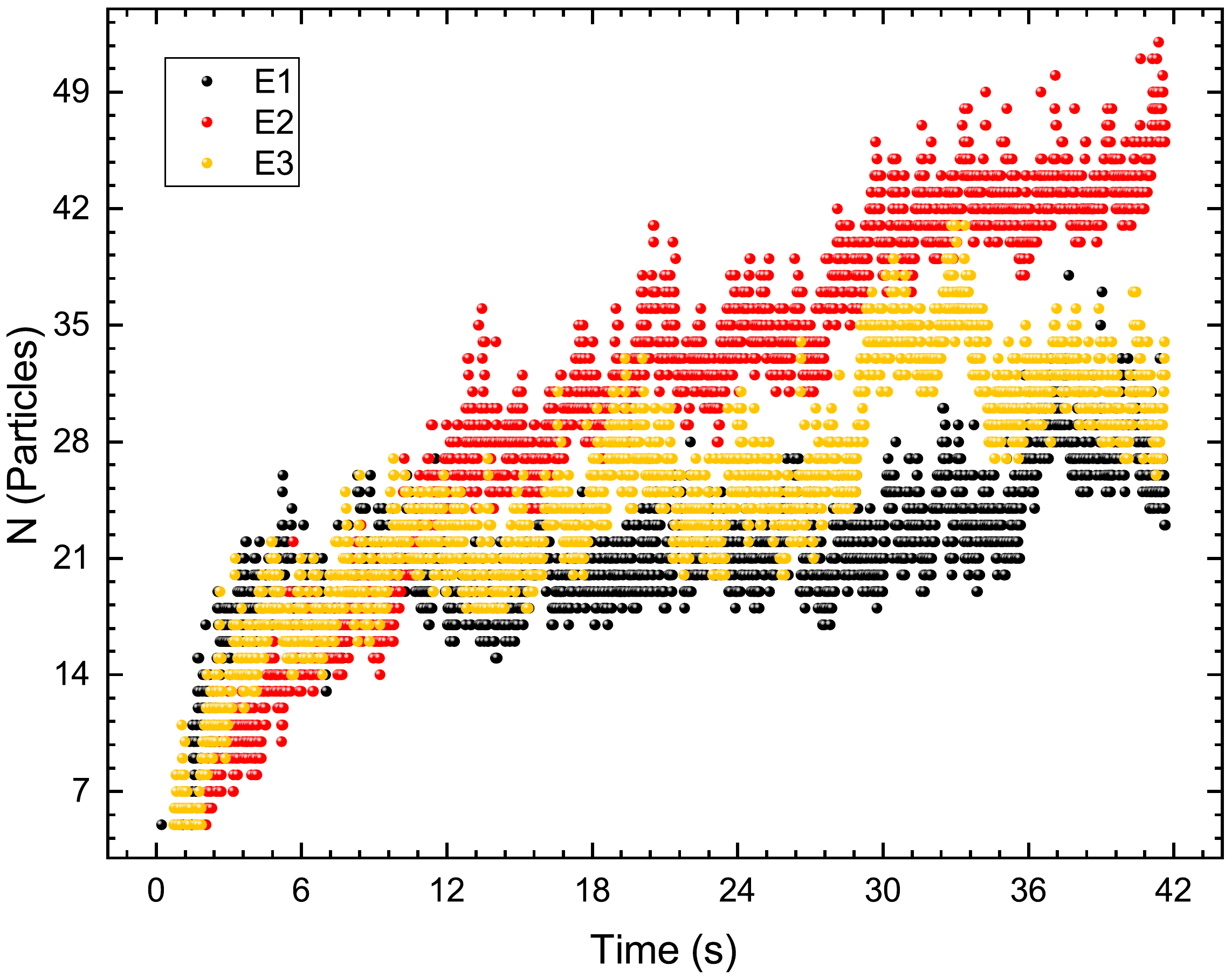}
\caption{\small  Comparison of the behavior of the aggregate size as a function of the time elapsed from its formation.  Recall the magnetic field amplitude decreases by 1~G in 50~s.}
\label{samestar}
\end{center}
\end{figure}
Although the formation and growing of a nucleus start at different magnetic field amplitudes between the three different experiments, the general evolution looks similar. In Fig. \ref{samestar} we compare the growing curve for each experiment using a temporal translation in such a way they start at the same point.  As discussed briefly in Sec.~\ref{systemaswhole}, the gaseous particles can diffuse roughly $\langle \Delta r^2 \rangle = 42\sigma^2$ in the 42~s period shown in Fig.~\ref{samestar}.

\begin{figure}
\centering
\includegraphics[width=0.45\textwidth]{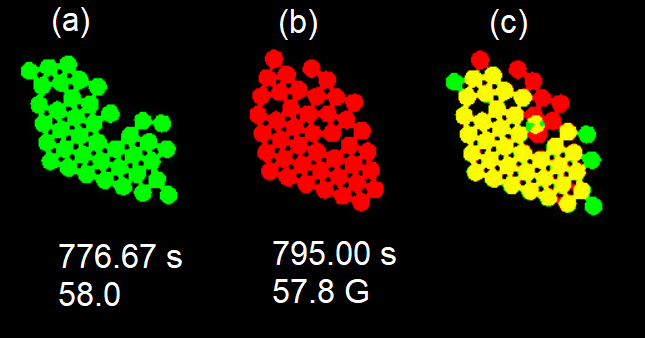}
\caption{ Comparison between the stable aggregate at two different times from experiment E1.  The growth suggests particles are added at concave regions where the new particles can contact more neighbors that help stabilize their presence in the aggregate, and particles with only tenuous connections to the aggregate are easier to ``evaporate.''}
\label{entropiceffect}
\end{figure}

The nucleus does not grow isotropically.  Gaseous particles randomly explore potential adhesion sites. Near the main stable aggregate, some small and unstable aggregate are formed all the time.  Often we observe that these small aggregates adhere to the stable aggregate, sometimes in an ordered way but more often in a disordered way. Due to this, aggregate does not grow in all directions at the same time, and the aggregate doesn't have a symmetric shape.  We observe that the most probable place to grow is where the aggregate presents a locally concave surface. Figure \ref{entropiceffect} shows an aggregate at two time points, showing that this aggregate grows by filling concave surfaces. It is also observed that particles with only one bond are more susceptible to be melted.  This supports our picture that frictional contacts of neighboring particles favor growth, and particles with few frictional contacts are easier to melt.

\section{Analysis of the birth and growth of a crystal}

At high temperature, particle motions are random and the system resembles a disordered gas.  Within this gas occasionally small aggregates form.  The particles have permanent magnetic dipoles, so frequently the early aggregates are in chain-like structures with the magnetic dipole moments aligned end-to-end.  Because of the many free particles rolling over the surface, aggregates experience frequent collisions which quickly dissolve these early aggregates.  As the magnetic field amplitude decreases and the system ``cools,'' small aggregates form more frequently and last for longer periods.  These are still usually chain-like unstable aggregates including dimers and trimers, and small ringlike structures. Less frequently, we observe the formation of bigger aggregates. 

As noted above, the stability of aggregates increases as the number of bonds increases.  Formation of ring-like and disk-like aggregates are less common than chain-like aggregates. However, once these more compact structures form, their stability is higher than a linear aggregate of the same size.  Additionally, the nearby gaseous particles produce an effective pressure toward the aggregate.  The effective pressure can overcome possible weak repulsive interactions between particles that can occur if their permanent magnetic dipole moments are oriented in a repulsive fashion.

As the system ``cools'' (lower oscillatory magnetic forcing), the kinetic energy of particles decreases, and a stable disordered aggregate forms.  The boundary of the aggregate fluctuates; particles rapidly join the boundary and other particles evaporate from the boundary.  Even particles remaining on the aggregate boundary rearrange due to kicks from the free particles.  Growth occurs when more particles join than leave the boundary.  Inside the aggregate, particles with four or more bonds occasionally rearrange to have six neighbors. We consider the crystalline nucleus is formed when we see a substructure with hexagonal ordering ($\psi_{6}^{'} \approx 1$) inside the stable aggregate. Thus the nucleus is an aggregate with at least one substructure with hexagonal order surrounding by particles a disordered configuration.

Further growth of the aggregate is reasonable to consider as a crystallization process.  We observe that for a while, the interior of the aggregate is ordered and the boundary is a bit more disordered; but after a certain size, the aggregate is completely ordered and further growth does not change this, as shown in Fig.~\ref{aggregatevstime}(b).

The aggregate grows or shrinks by the perimeter particles. A perimeter particle with only one neighbor is most susceptible to melting. A particle in the perimeter with a higher number of frictional contacts is more stable; locations where more frictional contacts are possible are thus good locations for a new particle to join the cluster.  This mechanism favors growth at concave regions of the aggregate surface. 

Figure \ref{comparpartbondphi} shows that as the aggregate grows, the mean number of neighbors $N_B$ increases.  Partially, this is due to a geometric effect:  boundary particles have fewer neighbors than interior particles, and a larger cluster has a larger ratio of interior particles to boundary particles.  That is, a cluster of size R has $\sim \pi R^2$ interior particles and $\sim 2 \pi R$ perimeter particles, so naturally the cluster average number of neighbors will get more and more dominated by the interior particles as $R$ increases.  Another cause of the increasing $N_B$ is the rearrangements of the interior particles, such as exemplified in Fig.~\ref{5and6}.  This mechanism is occurs because particles are in minimum energy positions when they are in a hexagonal lattice.

\subsection{Classical versus non-classical nucleation theory}

In the classical nucleation theory, the formation of a nucleus and the subsequent growth of the crystal phase is described as follows. The process begins with a supersaturated solution of the reagents. Spontaneous particle-concentration fluctuations drive the formation of a small ordered aggregate. This aggregate, depending on the balance of free energy and its size, could be more likely to shrink or grow depending on its size.  The critical size is defined as the size such that the probability of shrinking is equal to the probability of growth.  For aggregates larger than the critical size, the tendency is for the aggregate to continue growing and eventually forms a crystal.  Importantly, the structure of the critical nucleus is the same as the structure of the crystal.   

In contrast, in non-classical theories it is proposed that a nucleus can be formed from an amorphous aggregate of particles that eventually evolves into an ordered nucleus, that is, the formation of the nucleus occurs in at least two steps.  The concept is that the formation of an amorphous nucleus is easier than forming an ordered nucleus.  These proposals are supported by indirect evidence; it is challenging to obtain particle-level information about the nucleation process. 

We observe in our experiments that the initial process starts with an amorphous but quite stable aggregate, that both grows in size and becomes ordered over time.  The ordering generally starts in the interior of the aggregate.  Once the interior of the aggregate is hexagonally ordered, growth of the boundary of the aggregate nonetheless is still typically disordered.  Particles in a disordered configuration evolve into an ordered configuration by the kicks of the surrounding particles.  Thus we have experimental evidence supporting non-classical nucleation.  After a certain size, the crystal grows in an orderly process like  described by a classical nucleation theory.  This is generally a late stage of our experiment, where there are fewer gaseous particles.

\section{Conclusions and remarks}

We have studied the initial formation of the nucleus and the growing of a crystal. We have shown that in our system crystallization occurs according to non-classical nucleation theory in the early stages of the crystal growing and according to the classical description in the last stages. At the beginning small aggregates are formed because of particle concentration fluctuations. These aggregates are quickly destroyed by neighboring particles. As the temperature goes down, these aggregates lasted longer. At some moment an amorphous stable aggregate arose. Because the kicks of the neighboring free particles, the aggregate slowly becomes ordered, keeping approximately the same size. A crystalline nucleus arose inside this aggregate, a substructure with hexagonal order surrounded by still amorphous phase. The aggregate subsequently grows with a disordered boundary and further increased ordering within the interior, until eventually the entire aggregate is hexagonally ordered.  After that, all the aggregate is crystalline and further growing is according to the classical description.   Our work provides experimental evidence for a non-classical nucleation theory in the early stages of crystal growing.

\begin{acknowledgments}
The partial financial support by CONACyT, M\'exico, through grants 80629, and 731759 (Ciencia de Frontera) is acknowledged.  The work of E.R.W. was supported by the National Science Foundation under Grant No. CBET-1804186.
\end{acknowledgments}


\begin{thebibliography}{99}

\bibitem{ediger96} M.D. Ediger, C.A. Angell, S.R. Nagel, J. Phys. Chem. \textbf{100}, 13200 (1996).
\bibitem{debenedetti01} P.G. Debenedetti and F.H. Stillinger, Nature, \textbf{410}, 259 (2001).
\bibitem{stevenson2011} J. D. Stevenson and P.G. Wolynes, J. Phys. Chem. A  \textbf{115}, 3713 (2011).
\bibitem{sosso2016} G. C. Sosso, J. Chen, S. J. Cox, M. Fitzner, P. Pedevilla, A. Zen, and A. Michaelides, Chem. Rev.  \textbf{116}, 7078 (2016).
\bibitem{oxtoby2000} D. W. Oxtoby, Nature \textbf{406}, 464 (2000).
\bibitem{zahn2015} D. Zahn, Chem. Phys. Chem. \textbf{16}, 2069 (2015).
\bibitem{gebauer2014} D. Gebauer, M. Kellermeier, J.D Gale, L. Bergström and H. C\"olfen, Chem. Soc. Rev. \textbf{43}, 2348 (2014).
\bibitem{vekilov2010} P.G. Vekilov, Nanoscale \textbf{2}, 2346 (2010).
\bibitem{deyoreo2013} J. De Yoreo, Nat. Mater. \textbf{12}, 284 (2013).
\bibitem{gebauer2018} D. Gebauer, P. Raiteri, J. D.Gale and H. C\"olfen, Am. J. Sci. \textbf{318}, 969 (2018).
\bibitem{galkin1999} O. Galkin and P.G. Vekilov, J. Phys. Chem. B \textbf{103}, 10965 (1999).
\bibitem{yau2000}S.T. Yau and P. G. Vekilov, Nature \textbf{406} 494 (2000).
\bibitem{vekilov2011} P.G. Vekilov, Rev. Chem. Eng. \textbf{27}, 1 (2011).
\bibitem{gasser01} U. Gasser, E.R. Weeks, A. Schofield, P.N. Pusey and D.A. Weitz, Science \textbf{292}, 258 (2001).
\bibitem{konig2005} H. K\"onig, Europhys. Lett \textbf{71}, 838 (2005).
\bibitem{assoud2009} L. Assoud, F. Ebert, P. Keim, R. Messina, G. Maret and H. L\"owen, J. Phys.: Condens. Matter \textbf{21}, 464114 (2009).
\bibitem{wang2010} Z. Wang, A. M. Alsayed, A. G. Yodh, and Y. Han, J. Chem. Phys. \textbf{132}, 154501 (2010).
\bibitem{tan13} P. Tan, N. Xu, and L. Xu, Nature Phys. \textbf{10}, 73 (2013).
\bibitem{tsai2003} J.-C. Tsai, G. A. Voth, and J. P. Gollub, Phys. Rev. Lett. \textbf{91}, 064301 (2003).
\bibitem{rietz2018}F. Rietz, C. Radin, H.L. Swinney, and M. Schroter, Phys. Rev. Lett. \textbf{120}, 055701 (2018).
\bibitem{panaitescu10} A. Panaitescu and A. Kudrolli, Phys. Rev. E \textbf{81}, 060301(R) (2010).
\bibitem{panaitescu12}A. Panaitescu, K.A. Reddy and A. Kudrolli, Phys. Rev. Lett. \textbf{108}, 108001 (2012).
\bibitem{ebert2009} F. Ebert, P. Dillmann, G. Maret, and P. Keim, Rev. Sci. Instrum. \textbf{80}, 083902 (2009).
\bibitem{konig2004} H. K\"onig, K. Zahn and G. Maret, AIP Conf. Proc. \textbf{708}, 40 (2004)
\bibitem{sanchez19} M. J. S\'anchez-Miranda, J. L. Carrillo-Estrada and F. Donado, Scientific Reports \textbf{9} 3531 (2019). 
\bibitem{reis2006} P. M. Reis, R. A. Ingale, and M. D. Shattuck, Phys. Rev. Lett. \textbf{96}, 258001 (2006).
\bibitem{reis2007} P.M. Reis, R. A. Ingale, and M. D. Shattuck, Phys. Rev. E 75, 051311 (2007).
\bibitem{daniels2005} Karen E. Daniels and Robert P. Behringer, Phys. Rev. Lett. \textbf{94}, 168001 (2005).
\bibitem{voronoi18} R.E. Moctezuma, J.L. Arauz-Lara and F. Donado, Physica A \textbf{496} 27 (2018) .
\bibitem{crystal20} A. Escobar, C. Tapia-Ignacio, F. Donado, J.L. Arauz-Lara and R.E. Moctezuma, Phys. Rev. E \textbf{101}, 052907 (2020).
\bibitem{cafiero2000} R. Cafiero, S. Luding and H. J. Herrmann, Phys. Rev. Lett.  \textbf{84},  6014 (2000).
\bibitem{blair2003} D. L. Blair and A. Kudrolli, Phys. Rev. E, \textbf{67}, 021302 (2003).
\bibitem{wang2012} Z. Wang, F. Wang, Y. Peng, Z. Zheng, and Y. Han, Science \textbf{338}, 87-90 (2012).
\bibitem{morales2018} D.A. Morales-Barrera, G. Rodr\'iguez-Gattorno, and O. Carvente,  Phys. Rev. Lett. \textbf{121}, 074302 (2018).
\bibitem{brownian17} F. Donado, R. E. Moctezuma, L. L\'opez-Flores, M. Medina-Noyola and J. L. Arauz-Lara, Scientific Reports  \textbf{7} 12614 (2017).
\bibitem{cecilio16} C. Tapia-Ignacio, J. Garcia-Serrano and F. Donado, Phys. Rev. E \textbf{94}, 062902 (2016).
\bibitem{imagej}S.A. Schneider, W.S. Rasband, K.W. Eliceiri, Nature Methods {\bf 9}, 671 (2012). 
\bibitem{mosaic} I.F. Sbalzarini and P. Koumoutsakos, J. Struct. Biol. \textbf{151}, 182 (2005). 
\bibitem{tapia20} C. Tapia-Ignacio, R.E. Moctezuma, F. Donado, and E.R. Weeks, Phys. Rev. E \textbf{102}, 022902 (2020).


\end{thebibliography}
\end{document}